\documentclass{emulateapj}
\shorttitle{SZ EFFECTS FROM QUASARS} \shortauthors{LAPI, CAVALIERE
\& DE ZOTTI}
\begin{document}
\title{Sunyaev-Zel'dovich Effects from Quasars in Galaxies and Groups}
%
\author{A. Lapi\altaffilmark{1}, A. Cavaliere\altaffilmark{1}, and G. De Zotti\altaffilmark{2}}
\altaffiltext{1}{Astrofisica, Dip. Fisica, Universit\`a ``Tor
Vergata'', Via Ricerca Scientifica 1, I-00133 Roma, Italy}
\altaffiltext{2}{INAF, Osservatorio Astronomico di Padova, Vicolo
dell' Osservatorio 5, I-35122 Padova, Italy}
%
\begin{abstract}
The energy fed by active galactic nuclei to the surrounding
diffuse baryons changes their amount, temperature, and
distribution; so in groups and in member galaxies it affects the
X-ray luminosity and also the Sunyaev-Zel'dovich effect. Here we
compute how the latter is \emph{enhanced} by the transient
blastwave driven by an active quasar, and is \emph{depressed} when
the equilibrium is recovered with a depleted density. We constrain
such depressions and enhancements with the masses of relic black
holes in galaxies and the X-ray luminosities in groups. We discuss
how all these linked observables can tell the quasar contribution
to the thermal history of the baryons pervading galaxies and
groups.
\end{abstract}
%
\section{Introduction}
%
The thermal energy content of the hot interstellar (ISM) or the
intracluster medium (ICM) pervading galaxies or their groups and
clusters can be probed with the Sunyaev-Zel'dovich (1980, SZ)
effect. This arises when the hot electrons Compton upscatter some
of the CMB photons crossing the structure; then the black body
spectrum is tilted toward higher frequencies.

In the $\mu$wave band the tilt mimics a diminution $\Delta T_{\mu
w}\approx -5.5\, y$ K of the CMB temperature, proportional to the
Comptonization parameter $y\propto n\,T\,R$. This is evenly
contributed by the electron density $n$ and the temperature $T$;
in fact, what matters is the electron pressure $p=n\, k\, T$
integrated along a line of sight (see Fig.~1):
\begin{equation}
y = 2\, {\sigma_T\over m_e c^2} \int_0^R{d\ell\, p(r)}~.
\end{equation}

To now, SZ signals have been measured in many rich clusters at
levels $y\approx 10^{-4}$ or $\Delta T_{\mu w}\approx -0.5$ mK
(see Rephaeli 1995; Birkinshaw 1999; Zhang \& Wu 2000; Reese et
al. 2002). These levels are consistent with ICM temperatures $k
T\approx 5$ keV, sizes $R$ of a few Mpcs, and central densities
$n\approx 10^{-3}$ cm$^{-3}$. Similar values are indicated by the
standard cluster view based on gravitational potential wells of
virial depth $k T_v\propto G\, M/R$ dominated by the mass $M\sim
10^{15}\, M_{\odot}$ of the dark matter (DM). In such wells ICM
masses $m\approx 0.2\, M$ are settled in hydrostatic equilibrium
at specific energies $kT\approx kT_v$.

The above values also fit in with the X-ray luminosities
$L_X\propto n^2\, R^3\, \sqrt{T}\approx 10^{44}-10^{45}$ ergs
s$^{-1}$ emitted through thermal bremsstrahlung by the ICM.
Groups, on the other hand, are underluminous relative to clusters;
they emit far less than the baseline level $L_g\propto T_v^2$
scaled at constant $m\approx 0.2\, M$ after the DM rules. The
observed $L_X-T_v$ correlation is clearly steeper, and goes from
$L_X\propto T_v^3$ in clusters to $L_X\propto T_v^{4}$ or
$T_v^{5}$ in poor groups, albeit with a wide variance (O'Sullivan,
Ponman \& Collins 2003). In other words, the ICM in groups appears
to be underdense compared to clusters.

The origin of such lower densities is currently debated. One view
centers on extensive radiative cooling (Bryan 2000) which would
remove much low entropy gas. An alternative line of explanations
(see Cavaliere, Lapi \& Menci 2002, CLM02; and refs. therein)
focuses on the energy injections affecting the ICM equilibrium
while the DM is hierarchically accrued over dynamical timescales
$t_d$. The inputs are provided when the baryons in galaxies
condense to form stars possibly in starbursts, which then explode
as SNe; alternatively or correlatedly (Menci et al. 2003, in
prep.), the baryons accrete onto supermassive black holes (BHs)
energizing active galactic nuclei (AGNs). Such feedback actions
deplete the ICM density in the shallower potential wells by
causing from inside thermal outflow and dynamical blowout; they
also preheat the gas exterior to the newly forming structures, and
so hinder its inflow.

In any case, for groups in \emph{equilibrium} the values of $y$
can be anticipated from the continuum $L_X$ through the
model-independent relation (Cavaliere \& Menci 2001)
\begin{equation}
y/y_g = \left(L_X/L_{g}\right)^{1/2}\, \left(T/ T_v\right)^{3/4}~.
\end{equation}
Here $y_g\propto (1+z)^{3/2}\, T_v^{3/2}$ is the baseline value
scaled to the formation redshift $z$ (Cole \& Kaiser 1988). So for
groups where $L_X < L_g$ holds we expect \emph{depressed} $y$.

Are \emph{enhanced} SZ effects also possible, or even likely? What
can these tell about the processes affecting $n$ and $T$ in groups
and galaxies? Here we propose that a specific answer will come
from SZ observations.
%
\section{The transient regime}
%
We start from recasting $y_g\propto E/R^2$ in terms of the gas
thermal energy $E \propto p\, R^3 $ at equilibrium. A small group
or an early massive galaxy with their virial temperatures
$kT_v\approx 1$ or $0.5$ keV would produce SZ signals $\Delta
T_{\mu w}/0.5\, \mathrm{mK}\approx -5$ or $-3\times 10^{-2}\,
(1+z)^{3/2}$. Larger energies $\Delta E\ga E$ added to the ICM/ISM
are expected to enhance the SZ signals yielding $y/y_g\approx
\Delta E/E$.

Such may be the case with the AGNs (see Valageas \& Silk 1999; Wu,
Fabian \& Nulsen 2000; Roychowdhury \& Nath 2002) that produce
large total outputs, typically around $5\times 10^{61}$ ergs over
times around $10^{8}$ yr comparable to $t_d$ of their host
structures. Such outputs can drive a blastwave sweeping out the
gas mass and raising its pressure (see Platania et al. 2002; also
Yamada \& Fujita 2001).

%
\begin{figure}[t]
\epsscale{0.7}\plotone{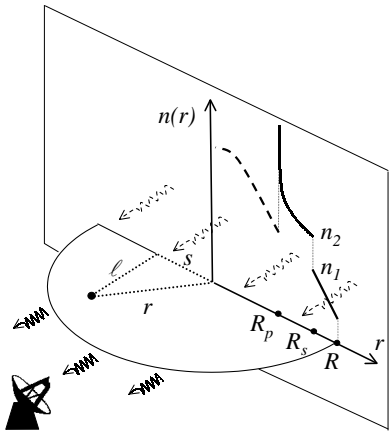}\caption{The geometry
underlying Eq.~(1). For a point in the structure $r$ is the radial
coordinate, $s$ is its projection on the plane of the sky, and
$\ell$ is the coordinate along the line of sight. On the vertical
axis we outline the initial density run, and the flow perturbed by
the AGN-driven blastwave.}
\end{figure}
%

We will see that the blast is \emph{constrained} by the coupling
of the AGN outputs to the surrounding gas, and \emph{restrained}
by the initial pressure and the DM gravity. We describe the blast
flow on using the self-similar hydrodynamical solutions presented
by CLM02. The simplest one obtains when the energy $\Delta
E(t)\propto t$ is delivered over times of order $t_d$ at the
center of an isothermal configuration for the DM and for the gas
with $n(r)\propto p(r)\propto r^{-2}$. Then the leading shock
moves out with uniform Mach number $\mathcal{M}$, i.e., with
radius $R_s = \dot{R}_s\, t$; the kinetic, the thermal and the
gravitational energies of the perturbed gas all scale like $R_s$.
So we can consistently define inside $R_s$ the total initial
energy $E$ (modulus); most important, our solutions provide
realistic predictions not only for \emph{strong} blasts but also
for the \emph{weak} ones driven by constrained values of $\Delta
E/E$.

Detailed profiles are presented in Fig.~2. Note that the perturbed
flow is confined to a shell; this is bounded by the leading shock
at $R_s$, and by a trailing contact surface (``piston'') located
at $\lambda\, R_s < R_s$ where the density diverges weakly while
the pressure is finite. So the relevant quantities may be also
obtained from the simple and precise ``shell approximation'' (see
Ostriker \& McKee 1988), which provides the energy balance in the
form
\begin{equation}
\Delta E - E = {1\over 2}\, m\, v_2^2 + {3\over 2}\, \langle p
\rangle\, V-{G\, M\, m\over R_s}~.
\end{equation}
Here $\langle p \rangle = p_2\, (5\,
\mathcal{M}^2+7)/(5\mathcal{M}^2-1)$ is the mean pressure within
the shell volume $V = 4\pi\, R_s^3\,(1-\lambda^3)/3$, important
for the SZ signals; $M$ and $m$ are the DM and the gas masses
within $R_s$. The Rankine-Hugoniot jump conditions yield the
postshock quantities: the velocity $v_2 = 3\, {\dot R_s}\,
(\mathcal{M}^2-1)/4\, \mathcal{M}^2$, the pressure $p_2=p_1\,
(5\mathcal{M}^2-1)/4$ and the density $n_2 = 4\, n_1\,
\mathcal{M}^2 /(\mathcal{M}^2+3)$, given the preshock values
$p_1$, $n_1$.

%
\begin{figure}[t]
\plotone{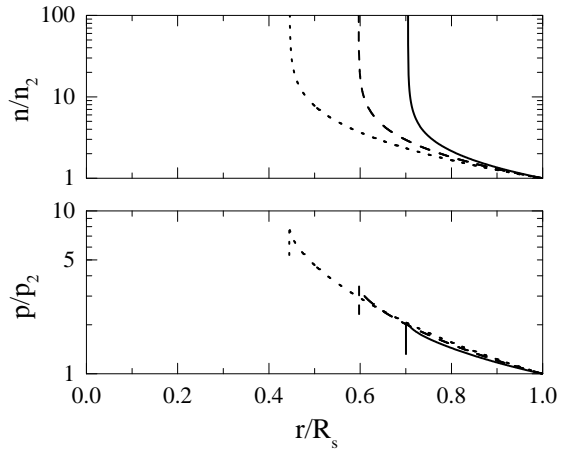}\caption{Radial distribution of density
and pressure in the blastwave, relative to the post-shock values
$n_2$ and $p_{2}$. {\it Solid lines} refer to a strong shock with
$\Delta E/E= 3$, {\it dashed lines} to an intermediate shock with
$\Delta E/E= 1$, and {\it dotted lines} to a weak shock with
$\Delta E/E= 0.3$; the corresponding piston positions are given by
$\lambda\approx 0.7$, $0.6$, $0.45$.}
\end{figure}
%

As to the energy $\Delta E$ actually injected over $t_d$, the
paradigm of supermassive BHs for the AGNs implies $\Delta E\approx
2\times 10^{62}\, f\, (M_{BH}/10^9\, M_{\odot})\, (1+z)^{-3/2}$
ergs when the mass $M_{BH}$ is accreted with conversion efficiency
of order $10^{-1}$. The fractional energy $f$ coupled to the
surrounding gas is poorly known. Including inefficiencies due to
low momentum transfer, non-spherical geometries and covering
factors it may range from $f\approx 10^{-2}$ for radio-quiet, to
some $10^{-1}$ for strongly absorbed (BAL) or radio-loud quasars,
a small minority. Average values $f\approx 5\times 10^{-2}$ are
shown below to be consistent with the observations both of the
relic BHs in galaxies and of $L_X$ in groups.

The ratio $\Delta E/E$ is seen from Eq.~(3) to be uniform, and to
constitute a key parameter for the shock strength. Table~1
presents quantities relevant to our computations of SZ signals.

We have extended our basic solution to initial density runs
$n\propto r^{-\omega}$, with $2\leq\omega < 2.5$. Then the initial
temperature declines as $T(r)\propto r^{2-\omega}$, and the shock
decelerates as $R_s\propto t^{2/\omega}$, again under
self-similarity; this also implies declining source luminosities
with $\Delta E(t)\propto t^{2\, (5-2\, \omega)/\omega}$. Table~1
shows that larger $\omega$ yields somewhat higher $\mathcal{M}$
and $\langle p \rangle$ at given $\Delta E/E$.

%
\section{Transient and equilibrium SZ effects}
%
In computing how $y$ is \emph{enhanced} during the blast transit,
we focus on $\bar{y} \propto R^{-2}\, \int {ds\, s\, y(s)}$
averaged over the structure area, which will subtend small angles
$R/D_A\la 1'$ for an early group or galaxy. Normalizing the shock
position as $x\equiv R_s(t)/R$, we find the full signal
\begin{equation}
{\bar{y}\over \bar{y}_{g}} = {\langle p\rangle\over 3\, p_1}\, (1
- \lambda^3)\, x + \sqrt{1-x^2}\simeq {\langle p\rangle\over 3\,
p_1}\, (1 - \lambda^3)~,
\end{equation}
in terms of $\bar{y}_g = (4\, \sigma_T/m_e c^2)\, p(R)\, R$. The
last approximation applies for $x \approx 1$, which maximizes the
transit time in the structure and optimizes the observability.

Strong SZ signals are seen from Eq.~(4) and Table~1 to require
substantial blasts driven through the ISM or the ICM, i.e., input
$\Delta E$ competing with the equilibrium value $E$. For
$\omega=2$ the latter writes $E\approx 2\times 10^{61}\,
(kT/\mathrm{keV})^{5/2}\, (1+z)^{-3/2}$ ergs; so the ratio reads
\begin{equation}
{\Delta E\over E} = 0.1\, {f\over 10^{-2}}\, {M_{BH}\over 10^9\,
M_{\odot}}\, \left({kT_v\over \mathrm{keV}}\right)^{-5/2}~.
\end{equation}
Here $M_{BH}$ is the mass accreted within $t_d\propto
(1+z)^{-3/2}$ by the central BH in a massive galaxy, or by the sum
of BHs shining within a group. Eq.~(5) yields $\Delta E$ close to
$E$ for a poor group with $kT_v\approx 1$ keV and $M_{BH}\approx
10^9\, M_{\odot}$. In going toward local, rich clusters $M_{BH}$
clearly lags behind $M$, so $\Delta E/E$ will decrease strongly,
see CLM02 for details.

At the other end, toward galaxies $\Delta E/E$ is constrained not
to exceed a few, lest the gas contained within kpcs and the
accretion it feeds are cut down (see Silk \& Rees 1998). The
pivotal value $\Delta E/E\approx 1$ recast in terms of the DM
velocity dispersion $\sigma = (kT_v/0.6\, m_p)^{1/2}$ reads
\begin{equation}
M_{BH} \approx 2\times 10^9 M_{\odot}\, \left({f\over
10^{-2}}\right)^{-1}\, \left({\sigma\over 300\, \mathrm{km\,
s}^{-1}}\right)^5~.
\end{equation}
Converting to the bulge dispersion $\sigma_{\star}\propto
\sigma^{1.2}$ (see Ferrarese 2002) yields $M_{BH}\propto
\sigma_{\star}^4$. For values $f \approx 5\times 10^{-2}$ the
relation accords with the observations in Tremaine et al. (2002).

%
\begin{deluxetable}{ccccc}
\tablecaption{Relevant Blastwave Quantities} \tablewidth{0pt}
\tablehead{\colhead{} &
\multicolumn{2}{c}{$\omega=2$} & \multicolumn{2}{c}{$\omega=2.4$}\\
\colhead{$\Delta E/E$} & \colhead{$\mathcal{M}$} &
\colhead{$\langle p \rangle/p_1$} & \colhead{$\mathcal{M}$} &
\colhead{$\langle p \rangle/p_1$}} \startdata
0.3 & 1.2 & 3.6  & 2.1 & 17.8\\
1   & 1.5 & 4.6  & 3.0 & 21.7\\
3   & 1.9 & 6.3  & 4.7 & 32.6\\
\enddata
\end{deluxetable}
%

Our results are represented in Fig.~3 vs. the depth $kT_v$ of the
host potential well. The square illustrates the \emph{minimal}
enhancement we expect from an early group at $z=1.5$ with $kT_v =
1$ keV, $f = 5\times 10^{-2}$ and $M_{BH}=10^9\, M_{\odot}$, so
with $\Delta E = 0.5\, E$. The bar gives a realistic upper
\emph{bound} for structures with steeper $n(r)$, namely, with
$\omega=2.4$; here $E$ is larger but the energy release is more
impulsive, resulting (see Table~1) in stronger signals. With radii
$R\approx 250$ kpc, the angular sizes $2\, R/D_A\approx 1'$ are
close to their minimum in the concordance cosmology (cf. Bennett
et al. 2003). Comparable resolutions will soon be achieved, see
\S~4.

The circles in Fig.~3 represent our results for a massive
($\sigma= 300$ km s$^{-1}$, $R\approx 100$ kpc) and still gas-rich
($m = 0.15\, M$) protogalaxy at $z=2.5$. The open circle refers to
$\Delta E=E$ or $M_{BH} = 6\times 10^8\, M_{\odot}$; the filled
one to $\Delta E=3\, E$ or $M_{BH} = 2\times 10^9\, M_{\odot}$,
just compatible with the scatter in the $M_{BH}-\sigma$
correlation. The related angular sizes are around $0.5'$; with
resolution fixed at $2\, \theta_{b}\approx 1'$, the signals will
be diluted after $(R/D_A\, \theta_b)^2\approx 1/4$ and scaled down
to $\Delta T_{\mu w}\approx -20\, \mu$K.

The inset represents the corresponding statistics. This is
evaluated on inserting the related blue luminosities $L = \Delta
E/10\, f\, t_d\approx 5\times 10^{45}$ and $1.5\times 10^{46}$
ergs s$^{-1}$ (with a bolometric correction $10$) in the quasar
luminosity function observed for $z\la 2.5$ by Boyle et al.
(2000), and discussed by Cavaliere \& Vittorini (2000, CV00). In
terms of the cumulative fraction of bright galaxies hosting a type
1 quasar brighter than $L$, this reads
\begin{equation}
N(L)\, L \approx 2\, 10^{-2}\, (1+z)^{3/2}\,
\left(L_b/L\right)^{2.2}~,
\end{equation}
beyond the break at $L_b = 5\times 10^{45}\, [(1+z)/3.5]^3$ ergs
s$^{-1}$. The same luminosity function interpreted in terms of
interactions of the host galaxy with its group companions (CV00)
yields a few signals per $10$ poor groups, with the strength
represented by the square in Fig.~3.

%
\begin{figure}
\plotone{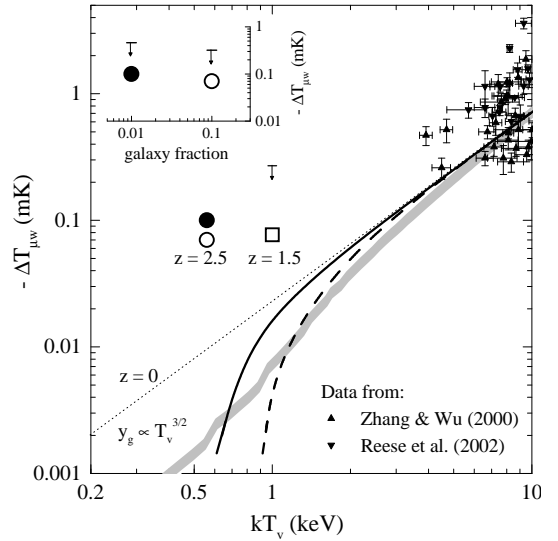}\caption{Predicted SZ signals as a
function of the virial temperatures of galaxies, groups and
clusters. {\it Dotted line}: the baseline $y_g$ at $z=0$. {\it
Shaded strip}: central signals from gas in the equilibrium set by
SN preheating (Cavaliere \& Menci 2001). {\it Thick lines}: same
for feedback from AGNs with $M_{BH}=10^9\, M_{\odot}$ and coupling
$f =3\times 10^{-2}$ ({\it solid}) or $f = 10^{-1}$ ({\it
dashed}). {\it Square}: area-averaged, undiluted signal from a
group at $z=1.5$, driven by AGN activity with $M_{BH}=10^9\,
M_{\odot}$ and $f=5\times 10^{-2}$; the bar represents the bound
for $\omega=2.4$. {\it Circles}: same from a massive galaxy at
$z=2.5$, for $\Delta E = E$ ({\it open}) or $3\, E$ ({\it
filled}); the inset specifies the statistics.}
\end{figure}
%

After the passage of the blast, the gas recovers hydrostatic
equilibrium. This may be described by $n(r)=n(R)\, \exp{(\beta\,
\Delta \phi)}$ for a nearly isothermal ICM in the (normalized) DM
potential well $\Delta \phi(r)$, see Cavaliere \& Fusco-Femiano
(1976). The blast heats up the gas so decreasing the value of the
parameter $\beta=T_v/T$, that we reset on using $T$ averaged over
the mass in the shell. The blast also ejects gas and depletes all
densities; we reset $n(R)$ by requiring the volume integral of
$n(r)$ to equal the gas mass left by the blast inside $R$ at
$t=t_d$.

The equilibrium SZ effect is then computed after Eq.~(1), and the
resulting signals are also plotted in Fig.~3. We recall from CLM02
that the equilibrium also provides good fits to the observed $L_X$
in groups for the \emph{same} coupling $f\approx 5\times 10^{-2}$,
consistent with Eq.~(6).

%
\section{Discussion and Conclusions}
%
This Letter centers on how the SZ effect is affected by the energy
fed back by AGNs into the surrounding gas. We predict both
transient, \emph{enhanced} and long-term, \emph{depressed} SZ
signals to be produced by an energy addition $\Delta E$ to the
equilibrium value $E$.

In $1$ keV groups the condition $\Delta E\la E$ holds, and the AGN
feedback has a considerable impact. A blast is driven through the
gas; during its transit the area-averaged SZ signal is enhanced as
the gas is just redistributed while its pressure is \emph{raised}.
But eventually a considerable gas fraction is \emph{ejected}; so
$n$ is depleted and $y\propto n\, T$ is decreased at equilibrium.
We have computed these effects under the \emph{restraints} set to
(weak) blast propagation by initial pressure and DM gravity.

The result on the SZ effect is twofold: on scales $1'$ we predict
\emph{transient} enhancements up to $\Delta T_{\mu w}\approx -80\,
\mu$K (a representative example at $z\approx 1.5$ is given in
Fig.~3), followed by \emph{long-term} depressions. The latter
correlate after Eq.~(2) with the equilibrium X-ray luminosities
$L_X\propto n^2\, \sqrt{T}$ which are very depressed.

Larger $\Delta E/E$, yet \emph{constrained} by Eq.~(6), yield
stronger SZ enhancements in gas-rich massive protogalaxies with
halo radii $R\sim 100$ kpc; representative examples are shown by
the circles in Fig.~3. With angular sizes $0.5'$, these may be
diluted down to $\Delta T_{\mu w}$ $\approx -20\, \mu$K when
observed at a resolution around $1'$.

Such resolutions will be achieved by several instruments now being
built or designed, enabling ``blind'' sky surveys for SZ signals
to $\mu$K sensitivities over tens of square degrees (see
Carlstrom, Holder \& Reese 2002). In particular, promising
perspectives are offered by multi-beam, high frequency radio
receivers like OCRA (Browne et al. 2000), and also by
interferometers equipped with wide-band correlators like ATCA, SZA
(Holder et al. 2000), AMI (Jones 2002), and AMiBA (Lo 2002). The
SZ signals we consider may contribute equally or more than
clusters to the excess power already detected at high multipoles
with BIMA (Dawson et al. 2002). In the (sub)millimetric band the
SZ signal is positive, and will be accessible to large bolometer
arrays like BOLOCAM, whose developments will enable deep, wide
surveys (Mauskopf et al. 2002). Eventually, ALMA
(\url{http://www.alma.nrao.edu/}) will provide in selected areas
higher resolution for both sides of the SZ effect.

Enhanced signals as discussed here would constitute
\emph{signatures} of strong feedback caught in the act. This is
specific of AGNs, since SNe feed back at most $0.3$ keV/particle
(see CLM02); on the other hand, extended cooling which depletes
$n$ without increasing $T$ hardly could enhance $y\propto n\, T$.
Interlopers might be introduced by merging events; however, these
primarily govern the growth of the DM halos and set the virial
$T_v\propto M^{2/3}$ included in our baseline $y_g\propto
T_v^{3/2}$. Only an exceptional major merging may contribute an
energy step sizeable but still bound by $\Delta E < E$. Even this
produces transonic inflows in the high-$T_v$ partner gas,
originating limited warmer features as picked up by highly
resolved X-ray studies of clusters. Still smoother inflows are
produced by SN preheating (see Voit et al. 2003), while stronger
blasts are driven by AGNs, in the galaxies and groups that we
propose here as primarily SZ objects.

Detecting $10$ such signals will require surveys over $500$
arcmin$^2$ at $1'$ resolution, based on the conservative surface
density of $10^2$ powerful quasars/deg$^2$ consistent with
Eq.~(7). For groups our evaluations (Fig.~3) lead to SZ
enhancements in the range $\bar{y}/\bar{y}_g\approx 1.2 - 4$ over
a depressed if wrinkled landscape. In fact, for $k T_v\approx 1$
keV the baseline $y_g$ is affected mainly by SNe (see Fig.~3);
these depress the average levels somewhat, and cause at $z\approx
1.5$ a $10\%$ relative scatter (at $96\%$ confidence). This
landscape may be sampled or bounded from independent groups
catalogued at comparable $z$. For massive protogalaxies
intrinsically stronger enhancements, less depression and narrower
scatter obtain. The candidate peaks are to be followed up with
ALMA for higher resolutions; in addition, optical $z$, and optical
$\sigma$ or X-ray $T$ will require current or moderately
extrapolated techniques (cf. Rosati, Borgani \& Norman 2002;
Shields et al. 2003).

In conclusion, we expect that AGN energy outputs around $10^{62}$
ergs with coupling $f \sim 5\times 10^{-2}$ leave two consistent
\emph{relics}: the depressed X-ray luminosities $L_X$ in local
galaxies and groups (see CLM02); the $M_{BH}-\sigma$ relation on
subgalactic scales (Eq.~6). Relatedly, on intermediate scales we
evaluate here (Fig.~3) \emph{transient} SZ signals standing out of
a generally depressed landscape. Such signals can provide real
time \emph{evidence} of AGN feedback acting on the diffuse baryons
in galaxies and groups. The evidence should be looked for
primarily in the SZ surveys that will be soon available.
\begin{acknowledgments}
We thank N. Menci and our referee for helpful comments.
\end{acknowledgments}

\end{document}